# Metasurface-encoded optical neural network wavefront sensing for high-speed adaptive optics

*Arturo Martin Jimenez[#1], Dylan Brancato[#1], Marc Baltes[1], Jackson Cornelius[1], Neset Aközbek[2], and Zachary Coppens[1*]*

[1]*CFD Research Corporation, Huntsville, AL, 35806*

[2]*US Army Space and Missile Defense Command, Huntsville, AL, 35808*

[*]*Email:* [zachary.coppens@cfd-research.com](zachary.coppens@cfd-research.com)

[#]*Authors contributed equally to this work*

**Abstract**

Free-space optical communications with moving targets, such as satellite terminals, demand ultrafast wavefront sensing and correction. This is typically addressed using a Shack–Hartmann sensor, which pairs a high-speed camera with a lenslet array, but such systems add significant cost, weight, and power demands. In this work, we present a hybrid opto-electric neural network (OENN) wavefront sensor that enables ultra-high-speed operation in a compact, low-cost system. Subwavelength diffractive metasurfaces efficiently encode the incoming wavefront into tailored irradiance patterns, which are then decoded by a lightweight multilayer perceptron (MLP). In simulation and experiment, the hybrid approach achieves average Strehl ratio (SR) improvements exceeding 60% and 45%, respectively, for unseen wavefronts compared to purely electronic sensors with few-pixel inputs. Although larger MLPs allow purely electronic sensors to match the hybrid's SR under static conditions, transient atmosphere modeling shows that their added latency leads to rapid SR degradation with increasing Greenwood frequency, while the hybrid system maintains performance. These results highlight the potential of hybrid OENN architectures to unlock scalable, high-bandwidth free-space communication systems and, more broadly, to advance optical technologies where real-time sensing is constrained by electronic latency.

**Introduction**

The rapid growth of data-intensive services over the past decade has driven an ever-increasing demand for higher bandwidth capacity. Optical fiber communication has risen to meet much of this demand, but it remains constrained by costly infrastructure and limited coverage. Satellite constellations with free-space optical links offer a compelling alternative, enabling global internet access without the burden of terrestrial infrastructure. Yet, while inter-satellite laser links now operate effectively, the satellite-to-ground segment still relies primarily on RF transmission. Shifting to bidirectional laser downlinks would unlock higher carrier frequencies, multi-gigabit data rates, spectrum license-free operation, and enhanced security. However, a main barrier to implementation is atmospheric turbulence, which distorts the optical wavefront, lowering coupling efficiency (CE) and raising bit-error rates (BER) [1–4].

Adaptive optics (AO) systems have long been used to correct turbulence effects in free-space optical communication links [3,5]. A typical AO system combines a wavefront sensor to measure distortions with a deformable mirror to apply corrective phases. For stable operation, the wavefront sensor must run at bandwidths multiple times faster than the Greenwood frequency ($f_G$), which scales with wind velocity. In satellite downlinks, effective wind velocity due to high slew rate during tracking can drive Greenwood frequencies up to ~1 kHz[6–9]. While Shack–Hartmann wavefront sensors (SHWS) can achieve temporal bandwidths near 40 kHz using lenslet arrays paired with high-speed Phantom cameras[10], these implementations are bulky, power-hungry, and costly. Real-time wavefront reconstruction from a SHWS also depends on slow, computationally intensive algorithms that often require specialized hardware like FPGAs or GPUs [11,12].

Optical neural networks (ONNs) shift the computation to the optical domain offering a means of reducing latency by eliminating electronic reconstruction altogether. Implemented as cascades of passive diffractive layers, ONNs map distorted wavefronts directly to irradiance patterns that can be captured by a few high-speed photodiode pixels, encoding aberration coefficients at the speed of light. This enables compact, ultrafast, low-power sensing without high-end electronics. While promising, ONNs require strict alignment tolerances (<$\lambda$/2) making their implementation at free-space optical communication wavelengths challenging[13–16]. These challenges can be overcome with photonic encoders that pair simpler (even single layer) optical front ends with lightweight artificial neural network (ANN) back ends for low-latency reconstruction. Demonstrations range from photonic lanterns paired with neural networks for Zernike mode recovery to hybrid optical-electronic classifiers for image datasets like MNIST[15,16]. While this encoder-based paradigm presents a promising way of balancing the strengths and weaknesses of optical computing and artificial intelligence, a detailed study of its efficacy for a complex problem such as wavefront sensing in atmospheric turbulence remains to be done.

In this work, we present an implementation of a hybrid opto-electric neural network (OENN) system for high-speed wavefront sensing in atmospheric turbulence. Our design employs two metasurfaces: a compact phase-diversity encoder that splits the incoming beam into two focal points with different defocus, and a diffractive ONN layer that converts each focused beam into encoded diffraction patterns. These irradiance patterns are then decoded by a lightweight multilayer perceptron (MLP) to generate the parameters required to drive a deformable mirror for correction. We developed an end-to-end optimization pipeline that jointly optimizes the metasurface encoder and MLP encoder, while incorporating the practical limitations of an actuated deformable mirror. We assessed the benefit of the hybrid system by comparing performance against an MLP-only wavefront sensor in metrics such as corrected Strehl ratio (SR) and computational latency. Across simulation and experiment, the hybrid sensor achieves average Strehl ratio improvements exceeding 60% and 45%, respectively, on unseen wavefronts relative to electronic-only systems with low-dimensional inputs. We also show that the hybrid system maintains high correction capability under transient atmospheric effects beyond 150 Hz, whereas the purely electronic systems with higher latency degrade rapidly and provide almost no improvement.

**Results**

Operational Concept

Our wavefront sensor (WS) design, shown schematically in Figure 1, is divided into two main components: the optical encoder and the backend digital decoder. These components work together to drive a DM used to correct incoming wavefront aberrations. The optical encoder itself is composed of two elements: a phase-diversity focusing metasurface (FMS) and a diffractive encoder metasurface which we denote as the single-layer ONN. The use of phase diversity for wavefront reconstruction has been shown to be highly effective for atmospheric turbulence, boasting high robustness even against high scintillation and the presence of branch points in the wavefront [17–19]. The FMS is the first to interact with the incident wavefront, focusing it onto the diffractive encoder metasurface at an axial distance of 50 mm. A supercell approach to the metasurface design provides focused light with two horizontally separated Point Spread Functions (PSFs) where one PSF is focused at the encoder plane and the other is focused 2.4 mm behind it. Each PSF is centered on a square metasurface (MS) of size 504 × 504 µm, forming two diffractive channels separated by a distance of 749 µm, as shown in Figure 1. The light diffracted from each channel propagates 2 mm to the detector plane where the intensity profiles are recorded.

The recorded relative intensities are used as the input parameters for a MLP whose output layer contains the values defining the stroke of the actuators in a DM. The linear superposition of the influence functions for all the actuators throughout the DM multiplied by their respective stroke value yield the sag of the mirror which can be converted to a phase profile that is added to the incident field. To counteract the effects of the atmosphere, the DM phase profile must be the conjugate of the incident wavefront. Since our sensor directly predicts DM stroke position for correcting the wavefront, it has lower latency than traditional AO systems that require computational reconstruction algorithms to convert measured phase to appropriate DM parameters.

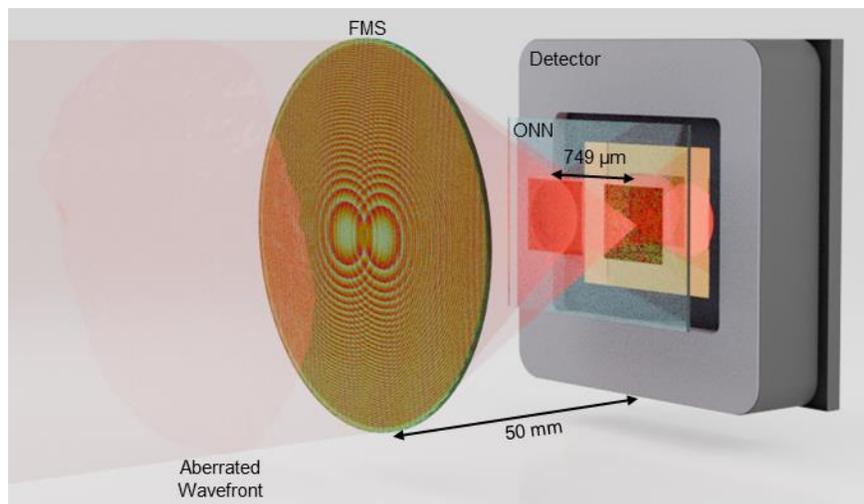

**Fig. 1 | Concept Figure.** Schematic of the optical encoder with the FMS and ONN MSs axially separated by 50 mm. The centers of the diffractive MSs on the ONN are separated by 749 µm. An incident beam whose wavefront has been distorted by the atmosphere, is focused by the FMS onto the ONN MSs, and is then diffracted into an intensity pattern captured by a few-pixel detector array.

Model Training

The unit cell parameters of the metasurface encoder and the model weights of the MLP were co-optimized through an end-to-end training pipeline to maximize performance (see Figure 2a). For the training data, a set of 10,000 input wavefronts were first generated by propagating a gaussian beam through a set of thirteen pseudo-random phase masks with power spectral densities defined by the Modified Von Kármán model for a nominal Rytov number of 0.1 [20]. The wavefronts were then added to the FMS phase and propagated 50 mm to create two horizontally shifted PSFs with phase diversity encoding on the ONN plane. The complex fields for these PSFs, along with the generated wavefront, were saved to be used for the training and testing of our models.

During training, the complex PSF fields were propagated through the metasurface ONN to a sensor plane where the captured irradiance values were pooled to accommodate various photodetector pixel sizes. The two channels in the ONN form two N × N arrays (where $N$ is the pool size) which are flattened into a $1 \times N^2$ array and passed to the MLP model. The MLP takes this input array and outputs an array of 199 values ranging from −1 to 1, corresponding to the normalized piston strokes of a deformable mirror (DM), with −1 (1) representing a full stroke in the negative (positive) direction, from a nominal (0) position. These values are used to calculate the correction phase that can be imparted by the DM, whose surface sag is modeled as the linear superposition of Modified Gaussian influence functions calculated for each actuator (see Supplementary 3 for more details) [21]. This correction phase is applied to the input field and is focused to obtain a PSF from which a Strehl ratio (SR) is calculated. In our simulations the corrected PSF is computed from the Fourier transform of the corrected field using Equation 1,

$$PSF(x',y') = \mathcal{F}\{U_c\} = \mathcal{F}\{U_i \exp(j\varphi_{DM}(x,y))\} \quad (1)$$

where $U_c$ is the corrected field, $U_i$ is the aberrated field incident on the FMS and $\varphi_{DM}(x,y)$ is the DM phase profile.

The SR is computed as the ratio of the PSF intensity at the origin ($x = 0, y = 0$) after correction to the peak intensity of a diffraction-limited PSF ($I_{max}$), with values ranging between 0 and 1. This final correction step simulates an AO configuration on the receiver side of an FSO link, where the incoming beam is corrected to increase coupling efficiency, or mixing efficiency in the case of homodyne detection. SR is easy to calculate and provides a suitable estimate of these metrics [5,22]. As such, we chose this as the figure of merit to be optimized during training, calculating the loss $L$ as, $L = 1-SR$. All training was done using Pytorch, which calculates the gradients for each computation step and automatically determines the appropriate change for each updatable parameter. Therefore, after the loss is calculated, both the MLP parameters and ONN phase profiles are updated. This process is repeated over several iterations for the entirety of the training set to minimize the loss.

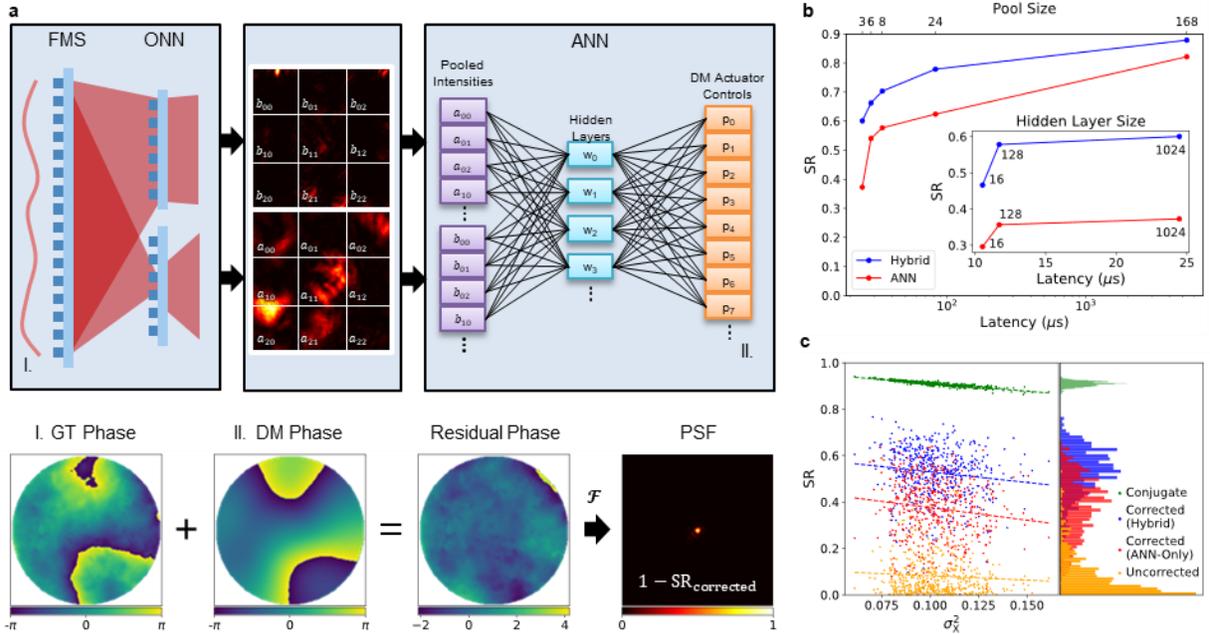

**Fig. 2 | Model training and simulated results. a**, Pipeline used to train/co-optimize ONN and ANN. The flowchart above shows the flow of information in the simulations, starting with modeling the propagation of the incident wavefront (example shown below I.) through the FMS and ONN, then the capture and of the intensity patterns on the detector plane and their integration over discrete square regions, and finally passing the digital intensity values through the MLP to obtain the DM parameters (II.). These parameters are then used to calculate the phase imparted by the DM and the residual phase after the correction can be used to calculate a PSF and corresponding SR. **b**, Results from the complexity study with varying input sizes (N = 3,6,8,24,168) for the hybrid (blue) and ANN-only (red) systems. The inset shows the results for the 3x3 systems for varying hidden layer sizes. **c**, SR values for the 3x3, 1x1024 models (hybrid in blue and ANN-only in red) plotted against Rytov number for each sample in the test set. Also shows uncorrected (orange) and conjugate (green) SR values for the same samples.

Simulated Results

The training pipeline described previously was used to perform a complexity study of the MLP, comparing the performance of the hybrid ONN-MLP system to that of an MLP-only system for various MLP sizes. Each training was performed using 9,000, 500, and 500 samples for the training, validation, and test, respectively. All models in this study were trained with the same hyperparameters for 150 epochs. For the first part of the study, only the input layer size was varied, simulating a reduction in detector resolution. Lower input sizes not only decrease the latency of MLP operation, but also enables the use of few-pixel photodiode arrays which can operate at much higher frame rates than focal plane arrays (FPAs). For the second part of the study, the hidden layer size was varied for a given input layer size. This change does not affect the latency of the sensor to capture the ONN signal and instead only impacts the performance and latency of the MLP model. The output layer size was fixed to 199 for all our tests. This corresponds to the number of actuators on the modeled DM, which constitutes a constraint on the spatial resolution for wavefront correction, and consequently the maximum achievable SR. For this number of actuators and a maximum stroke of ±1.5 μm, we found a maximum average SR value of 0.88 for wavefronts with nominal Rytov number of 0.1.

Figure 2b shows the results obtained for various pooling sizes (N = 3,6,8,24,168) and a single hidden layer of size 1024 with the hybrid ONN-MLP system and the MLP-only system. The hybrid system can be seen to outperform the purely electronic system for all input sizes with both models trending to converge to a common point at increased sizes. The advantage of the hybrid system is most significant at the lowest sizes, with a roughly 60% increase in average SR compared to the MLP-only system at 3x3 pooling. At this pooling size, the hybrid system enables a 240% reduction in latency for only a 23% reduction in SR when moving from the 24x24 to the 3x3 configuration. This could be a fair trade for many FSO systems where AO bandwidth is a high priority. Further decrease of the input dimensions yielded more reductions in latency but with more significant decrease in Strehl ratio; therefore, we chose the 3x3 configuration for further study. The inset plot in Figure 2b shows the results with a 3x3 input size and hidden layer sizes of 16,128, and 1024, for the two systems. As in the case for varying pooling sizes, the hybrid system achieves higher average SR than the purely electronic system for all hidden layer sizes. Further latency reduction can be appreciated by decreasing the hidden layer size, but a similar diminishing performance is obtained for sizes below 128.

To evaluate the performance of the hybrid system with 3x3 input size and 1024 hidden layer size, the SR calculated for the 500 samples from the test set are plotted in Figure 2c. The x-axis shows the Rytov number calculated for the corresponding input wavefront of each sample. The conjugate SR is that obtained from perfect phase correction (i.e., 0 residual phase error). It is important to note a general trend of decreasing average SR with increasing Rytov number, even for the conjugate case, which is a consequence of the phase-only correction. The uncorrected SR is calculated from the input wavefronts without imparting any phase correction. This achieves an average SR below 0.1 over all samples, while the corrected wavefronts achieve a mean SR of 0.60 when using the hybrid model, 1.6x higher than the 0.37 obtained when using the purely electronic model. As suggested by the results of the complexity study, the performance can be improved to approach the limit imposed by the DM if the model complexity is increased, at the expense of increased latency. However, increased latency also reduces the maximum SR achievable when the temporal behavior of atmospheric turbulence is considered, as we will demonstrate later in the experimental section.

Metasurface Design and Fabrication

The optimized ONN design for the 3x3 configuration was selected for fabrication. The metasurfaces used in this work, shown in Fig. a-b, were fabricated using electron-beam lithography followed by reactive-ion etching (see Methods). In addition to the FMS and ONN, auxiliary holograms were employed to facilitate precise alignment of the metasurfaces within the wavefront sensor. Rotating double-helix phase masks were used for the longitudinal alignment holograms (LAH) and the focus alignment hologram (FAH). These phase masks create point spread functions (PSFs) with two lobes that rotate with defocus or separation between the FMS and ONN [23,24]. The transverse alignment holograms (TAHs) are based on a design presented by Ghahremani et al. (2024) who demonstrated its use for sub-nanometer alignment precision [25]. See Supplementary 1 for more information on the design of these alignment holograms.

The insets in Figures 3a and 3b show scanning electron microscope (SEM) images of the FMS and ONN, respectively. For the ONN image, we can observe square areas of repeated pillar widths. This is a result of the sub-sampling used during the optimization of the ONN phase profiles, which was necessary to reduce training times and to reduce the probability of non-global convergence. All metasurfaces comprise the same unit cell material structure of Si pillars on an $SiO_2$ substrate. The design and modeling of these structures was performed using rigorous coupled-wave analysis (RCWA) to obtain the phase and transmission of a unit cell, shown schematically in the inset of Figure 3c. The curves on Figure 3c show the phase (solid line and circles) and transmission (dashed line and triangles) for various pillar widths ranging from 150 nm and 405 nm and a fixed pillar height of 1200 nm with a unit cell periodicity of 700 nm. Pillar widths from 325 nm to 345 nm are omitted to avoid a resonance. The discrete dots (blue circles for phase and orange triangles for transmission) denote the 16 selected pillar widths that were used for fabrication of the metasurfaces, corresponding to evenly-distributed phase values between 0 and $2\pi$. Matching these phase values to the phases at each location of a given metasurface design creates a map of pillar widths that can be used to lithographically pattern the metasurfaces (see Methods).

The final phase profiles for the FMS and ONN are shown in Figure 3d-e. The FMS uses a supercell scheme to combine two hyperbolic phase profiles in one aperture (see Methods Supplementary 2). This scheme consists in creating a checkerboard pattern as shown in the inset at the bottom left corner of Fig. 3d where every other pixel value corresponds to the same hyperbolic profile. The composite phase obtained is designed to focus an incoming collimated beam to two horizontally displaced points separated by 749 µm. One point is focused at an axial distance of 50 mm while the other one focuses at 52.4 mm. With the ONN placed at 50 mm from the FMS, this creates a 'focused' and 'defocused' PSF on the ONN plane, as shown in the inset at the bottom right corner of Fig. 3d. As such, the metasurfaces on the ONN are horizontally separated by 749 µm and centered with respect to these PSFs. The degree of defocus for the second PSF ensures there is an appreciable difference between the inputs to each ONN and provides sufficient phase diversity.

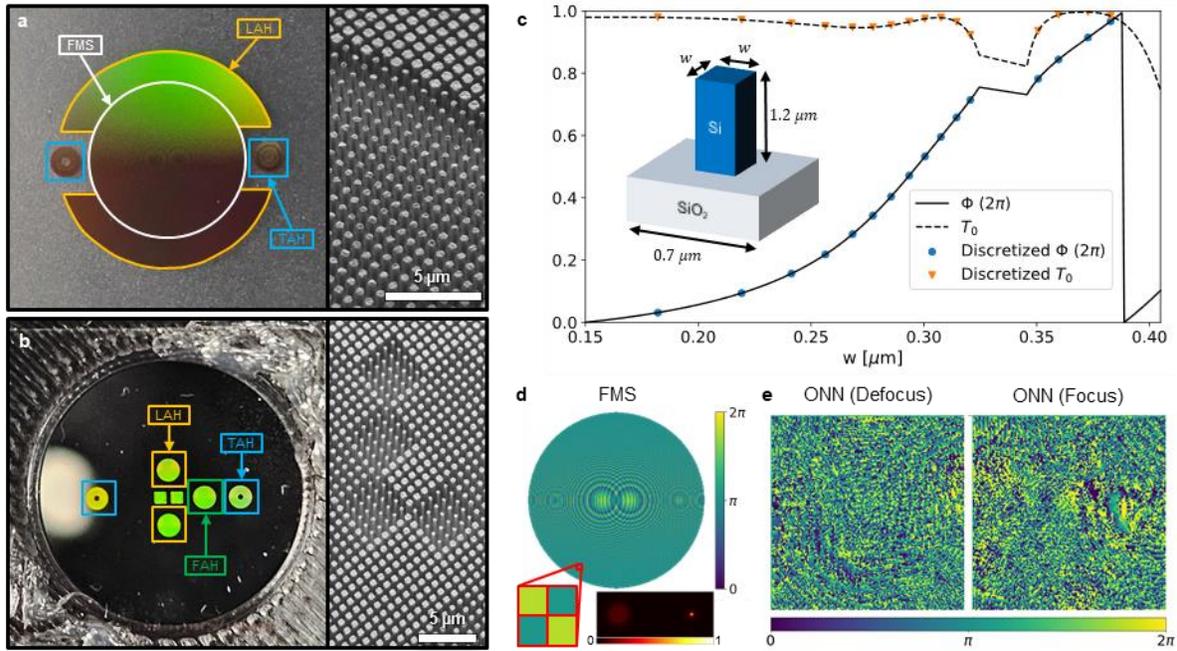

**Fig. 3 | Metasurface Design and Fabrication. a**, Image of the fabricated FMS with labeled LAH (orange) and TAH (light blue) and inset on the right showing an SEM image of a portion of the MS. **b**, Image of the fabricated ONN with LAH (orange), TAH (light blue) and FAH (green) and inset on the right showing an SEM image of a portion of the ONN MS. **c**, Transmission for diffractive $0^{th}$-order and phase for Si on $SiO_2$ nanostructures with pillar widths from 0.150 µm to 0.405 µm obtained from RCWA. The inset shows a schematic of the unit cell with a period of 0.7 µm and a pillar height of 1.2 µm. Discretized pillar widths selected for fabrication are denoted by the orange triangles (transmission) and the blue dots (phase). **d**, Phase profile for FMS with bottom left inset showing the checker-board pattern of the super-cell scheme and the bottom right inset showing the PSFs obtained from a flat wavefront. **e**, Phase profile for fabricated ONN MSs whose design was obtained from optimization of the 3x3 1x1024 hybrid system.

Experimental Results

To demonstrate the feasibility of the hybrid system for wavefront measurements, the fabricated FMS and ONN were placed in the experimental setup shown in Fig. 4a. We used a 1550 nm fiber-coupled laser as the illumination source, which is properly collimated and expanded to illuminate a liquid crystal spatial light modulator (SLM) (Meadowlark's 1920x1200 HDMI SLM). The SLM is used to impart a series of phases corresponding to turbulent wavefronts generated in the same manner as those used for training our models in simulation. The modulated laser reflects off of the SLM and propagates towards our wavefront sensor, whose optical setup comprises the FMS and ONN in proper alignment with each other and the imaging optics. This alignment was achieved, as mentioned in the previous section, through the use of holographic phase masks co-planarly fabricated with the FMS and ONN. A relay lens was used to project the image created by the ONN onto the focal plane of a SWIR camera (Goldeye G-033). For each SLM phase, a 14-bit capture of the intensity pattern created by the ONN averaged over 20 frames was saved. 50,000 samples were saved in this way with the ONN in the system, then 50,000 samples were saved using the same SLM phases with the ONN removed from the system.

In Fig. 4b, we show an example experimental capture and phase reconstruction on a wavefront generated by the SLM. The top row includes ONN results while the bottom row shows results without the ONN. All the intensity captures qualitatively match the simulated intensity distribution, giving confidence in the metasurface fabrication and experimental setup. The predicted phases were obtained after training MLPs with 2 hidden layers of size 4096 on the datasets. The model size was increased compared to simulation so that comparable average SRs could be achieved. We attribute the need for a larger model to several factors with the experimental data, such as noise, quantization, and limited dynamic range of the camera, which add additional complexity to the data, making it harder to fit. Nonetheless, experimentally comparing the performance of the hybrid and purely electronic system demonstrate the advantage of the ONN. Fig. 4c shows four boxplots with the distribution of SR values obtained from test set inference using two different models. The "Hybrid" models were trained on the data captured with the ONN, while the "ANN-only" models were trained on the data captured without the ONN. All hyperparameters were maintained during each model training to provide a fair comparison. It is evident that the hybrid models consistently achieve higher SR values than the ANN-only models, even at the larger model sizes of 2x4096 which show a 45% increase in SR. Interestingly, the 2x4096 ANN-only model performs approximately as well as the 1x1024 Hybrid model, under static turbulence conditions. The added latency from the larger model size, however, becomes a significant inhibitor when the transient effects of turbulence are considered.

During the training of our models, the atmosphere is simulated in a "frozen" state, and each sample corresponds to a random independent instance of a different atmosphere with the same nominal Rytov number. However, to gauge the true performance of each system it is also required to consider the temporal behavior of the atmosphere. Using the Greenwood frequency ($f_G$) as a measure of the turnover rate of the atmosphere (i.e., how fast the atmospheric aberrations change within the beam path), we can estimate its effect on the SR using Equation 2 [5,9,22] (see Methods). Here, the SR value is scaled by an exponential term that depends on the residual wavefront error $\sigma_t$ which is given by Equation 3 and depends on the ratio of $f_G$ to the closed-loop bandwidth ($f_{3dB}$) of the AO system. For our calculations, we took $f_{3dB} = 0.1/t_{WFS}$, where $t_{WFS}$ is the latency of our model and the 0.1 factor comes from following the 10x rule of thumb commonly used for the sampling rate of an AO system [26–28].

$$SR_t = SR e^{-\sigma_t} \tag{2}$$

$$\sigma_t = \left(\frac{f_G}{f_{3dB}}\right)^{5/3} \tag{3}$$

Fig. 4d shows the resulting scaled average SR for the hybrid 1x1024 model vs. the purely electronic (ANN-only) 2x4096 model as a function of Greenwood frequency. The hybrid model has a latency of 24.5 μs while the ANN-only model has a latency of 1480 μs. While both models start at roughly the same value ($f_G = 0\ Hz$), the performance of the ANN-only model quickly degrades for increasing $f_G$ values, while the performance of the hybrid model is largely unaffected. This demonstrates the advantage of the hybrid system, enabling faster wavefront reconstruction to retain AO benefits in applications with rapidly changing atmospheres, such as satellite-to-ground FSO links.

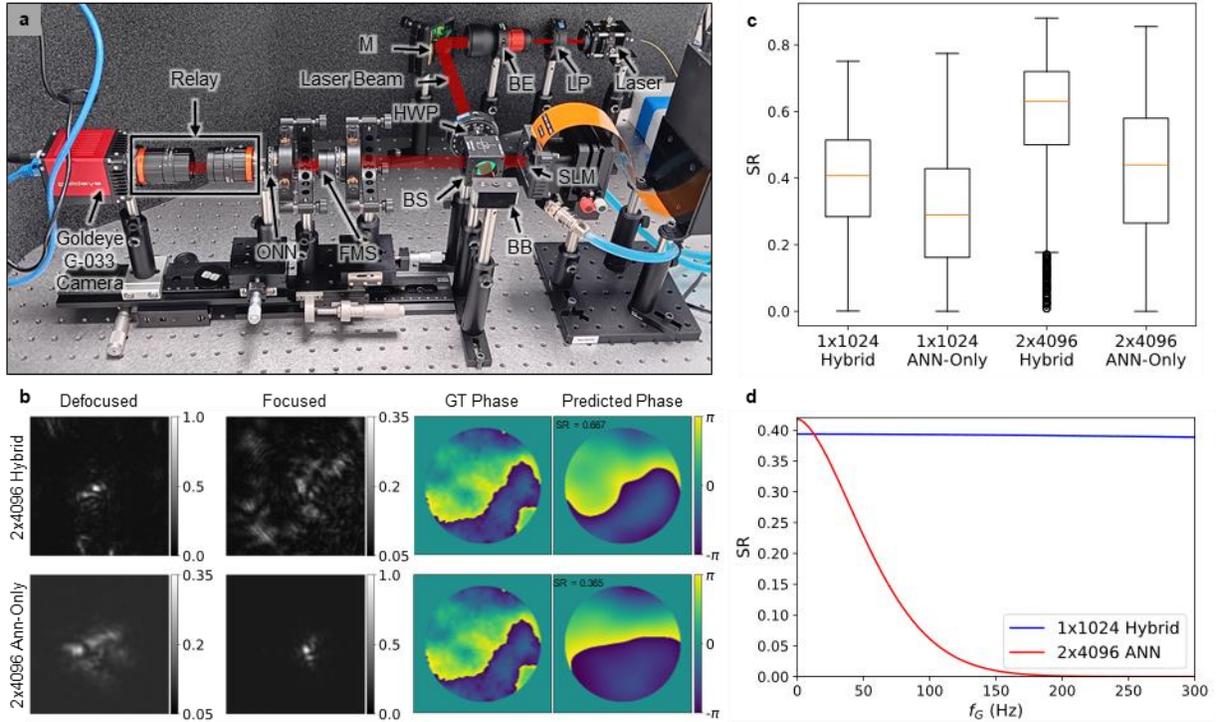

**Fig. 4 | Experimental Demonstration. a**, Experimental setup using a 1550 nm fiber-coupled laser, collimated and expanded, to illuminate an SLM, which imparts a random phase and reflects the beam towards our wavefront sensor comprised of the FMS, ONN, relay optics and the Goldeye G-033 SWIR camera. **b**, Experimental intensity captures from defocused and focused channels are shown in the first and second column for the setup with (top row) and without (bottom row) the ONN MSs. The ground truth (GT) and predicted phases (obtained with a 2x4096 MLP) for each are shown in the third and fourth column with the calculated SR values. **c**, Boxplots constructed from the SR values calculated for 2,500 samples processed by the hybrid and ANN-only models with two MLP sizes (1x1024 and 2x4096). The vertical lengths of the boxes represent the extent from the first (Q1) to the third quartile (Q3) of the data, and the orange line marks the median. The whiskers extend from the box to the farthest data point lying within 1.5× the interquartile range from the box and the black circles are outliers. **d**, SR values for the 1x1024 hybrid (blue) and 2x4096 ANN-only (red) systems with increasing $f_G$ considering the measured latencies of each system.

## Discussion:

Our proposed hybrid ONN-ANN architecture shows great promise for ultra-high-speed wavefront sensing. There are, however, several areas for improvement as we push toward practical implementations of the technology. Here, we discuss a few considerations needed to bridge the gap as the focus for future work.

Increasing the physical fidelity of our simulations can improve the optimization of the hybrid (ONN + MLP) system. Factors such as noise, signal digitization, and the limited dynamic range of our detectors need to be considered to determine realistic performance limitations with real data. This is evident when comparing the performance achieved with experimental data vs. simulated data in this study. Adding these effects to simulation will allow the system to optimize both the optical encoder and MLP decoder under these non-ideal conditions, which has been shown to produce noise tolerant wavefront sensors[32].

In our experiment, we used a liquid crystal SLM to simulate the effects of atmospheric turbulence. While this allowed us to generate many independent instances of turbulence, which was necessary for training our models, it places a limit on the atmospheric temporal behavior that can be properly simulated. Additionally, we were limited by the frame rate of our camera when capturing the signal from the ONN. To properly demonstrate the high-speed capabilities of our approach, we could use rotating phase screens to generate time-varying atmospheric turbulence as well as multi-pixel detector arrays to capture the signal from the ONN at much higher frame rates than a conventional camera allows. Additionally, we could improve our model's reconstruction accuracy while maintaining its fast performance by implementing a larger MLP on a GPU. While the latencies reported here were all measured on a CPU (see Methods for more details) to maintain consistency, we note that using a GPU could enable the use of larger models (like the 2x4096) for practical systems. We have shown that pairing these models with an ONN achieves higher SR than can be obtained with just the MLP.

We have demonstrated the use of a hybrid ONN-ANN system for high-speed wavefront sensing in atmospheric turbulence. The advantages of a hybrid model over a purely electronic one are shown, with a 60% increase in average SR for the case of static turbulence in simulation (45% in experiment) with no additional latency. Our experimental results also reiterate the superiority of the hybrid model for all MLP complexities used, emphasizing its stability of reconstruction accuracy for Greenwood frequencies beyond 300 Hz. The hybrid system allows for the use of ultra-light MLPs achieving substantial reductions in wavefront sensing latency and improving the correction ability of AO systems. This system can be implemented with cheap photodiode arrays and lightweight electronics, for an ultra-low-SWAP wavefront sensor. It is also easily scalable to achieve higher reconstruction accuracy, while still maintaining lower latencies than conventional sensors. Further development of this method could yield state-of-the-art performance at unprecedented operational speeds and potentially extend its effectiveness to the deep turbulence regime.

**Methods**

Metasurface fabrication

For the fabrication of the metasurface, the material system chosen was silicon on fused silica, as it provides good transmission for the design wavelength and can be etched using the Bosh process. An OASIS file was created containing the widths of each resonator at their corresponding positions on the metasurface. For the final design, a unit cell period of 0.7 µm was used, with a Si pillar height of 1.2 µm and pillar widths ranging from 0.1 µm to 0.5 µm. Plasma-enhanced chemical vapor deposition (PECVD) was used to create a 1.2-µm-thick layer of silicon on a 500 µm fused silica wafer. Then, electron-beam lithography was used to pattern the silicon layer followed by a deep reactive-ion etching to create the pillar structures.

Numerical Wave Propagation

All optical wave propagations were done in Python using the Angular Spectrum Method [29] with FFT. Using this method, the propagated field is related to the initial field as shown in Equation 2.

$$U(x, y, z) = \mathcal{F}^{-1}\{M\mathcal{F}\{U(x, y, 0)\}\} \qquad (4)$$

$$M = \begin{cases} e^{i2\pi z \sqrt{\left(\frac{n}{\lambda}\right)^2 - (f_x^2 + f_y^2)}}, & \text{for } f_x^2 + f_y^2 \leq \left(\frac{n}{\lambda}\right)^2 \\ 0, & \text{otherwise} \end{cases}$$

This method was used to simulate the optical propagation through our FMS and ONN, modeling each metasurface as a phase mask. Additionally, it was used in the wavefront generation to propagate the point source field through the discretized turbulent atmosphere using split-step propagation. Split-step propagation assumes the effects of propagating the field through a volume with nonuniform refractive index can be captured by doing multiple discrete propagations and adding at each step the accumulated effect of the distance propagated.

Atmospheric Temporal Behavior and Effects of Model Latency on Strehl Ratio

For training, all data was generated using random independent instances of the atmosphere with the same nominal parameters such as $C_n^2$, propagation distance, inner and outer scales, etc. This accurately captured the spatial behavior of atmospheric turbulence, but the temporal behavior was assumed to be static during the time scale that the sensing occurred. This assumption, however, is not always valid, and leads to an overestimation of the SR achieved after correction. Considering the temporal rate of change of the turbulence becomes increasingly important when the latency of the AO system is high. This rate of change can be described by the Greenwood frequency ($f_G$) which is defined by the expression:

$$f_G = \left[0.102 \left(\frac{2\pi}{\lambda}\right)^2 \int_0^L C_n^2(z) V^{5/3}(z) dz\right]^{3/5} \quad (5)$$

Where $\lambda$ is the wavelength of light, $L$ is the propagation distance, $C_n^2$ is the refractive index structure parameter and $V$ is the transverse wind speed. Both $C_n^2$ and $V$ can vary along the propagation distance and often additional models, such as the Hufnagel-Valley for $C_n^2$ and the Bufton for wind speed, are used to describe them [9,30].

To estimate the effects of the AO correction on the SR, we use [31]:

$$SR \approx \exp(-\sigma_\phi^2) \quad (6)$$
$$\sigma_\phi^2 = \sigma_{WFS}^2 + \sigma_{DM}^2 + \sigma_t^2 \quad (7)$$

Where, $\sigma_{WFS}^2$ is the residual phase error from the wavefront sensor measurement, $\sigma_{DM}^2$ is the residual phase error from the DM correction, and $\sigma_t^2$ is the residual phase error associated with the temporal behavior of the AO. If we assume the SR calculated from our model ($SR_0$) already includes the first two error terms, we can estimate the effects of the temporal behavior as:

$$SR \approx \exp(-\sigma_{WFS}^2 - \sigma_{DM}^2 - \sigma_t^2) = SR_0 \exp(-\sigma_t^2) \quad (8)$$

Which is equivalent to the expression in Equation 4 and $\sigma_t^2$ is given by Equation 5.

Latency Measurements

All latencies reported here are measurements of the forward pass runtimes of the MLP models in C++ on the CPU averaged over 1000 runs. The models were originally written in PyTorch. We used the LibTorch library to serialize the models so they could be loaded and ran in C++ to capture it via compilation and better estimate their execution latencies. The latencies reported for the models used on the simulated data were measured on an AMD Ryzen Threadripper PRO 5955WX CPU with 16 cores. The latencies reported for the models used on the experimental data were measured on an AMD Ryzen Threadripper PRO 7955WX CPU, also with 16 cores. Measurements on an NVIDIA RTX A6000 Ada Generation GPU were also made for all models. The results can be seen in Supplementary 4.

# Supplementary

**Contents**

# S1 Alignment Holograms

Due to the high phase gradients imposed by the optical neural network (ONN), even minor misalignments can lead to significant deviations in the sensor plane output [1]. Due to the lack of access to a metrology setup, we employed low-cost alternative for alignment aid: a series of alignment holograms, each producing a distinct point spread function (PSF). Furthermore, the holograms enable a step-by-step alignment procedure that can significantly decrease the misalignment error. The spatial arrangement of these holograms relative to the ONN and focusing metasurface are illustrated in Supplementary Figure 1 below.

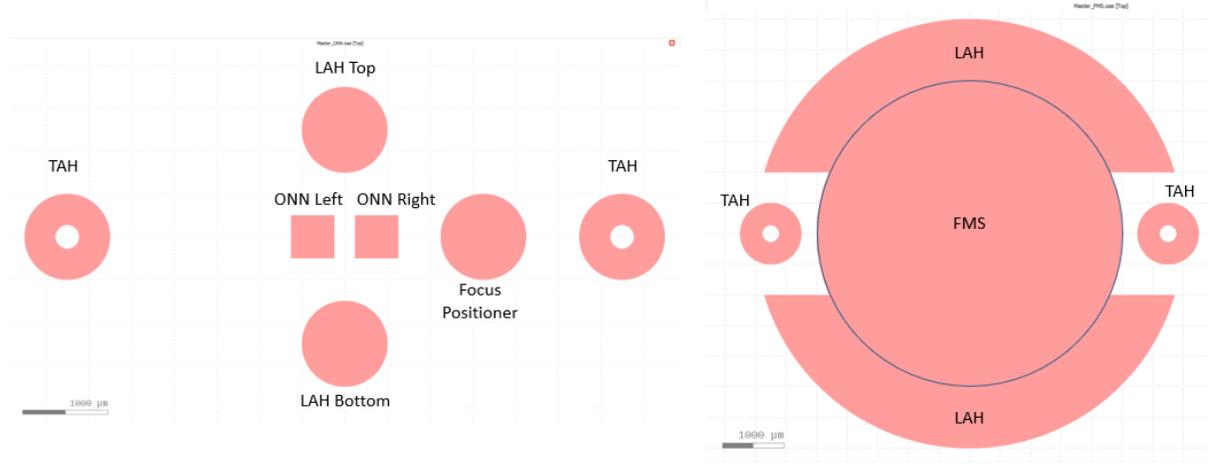

**Supplementary Fig. 1 | Encoder and FMS Schematic.** Schematic of the Encoder (ONN) and FMS files used to create the metasurface on a silicon wafer. The alignment holograms for each wafer are shown in reference to the sensing portions of the wavefront sensor.

## S1.1 Focus Positioner Hologram

The alignment holograms are designed to correct for tip, tilt, defocus, rotation, and transverse/longitudinal misalignments. The alignment process begins with the placement of the ONN wafer into the imaging system, which consists of a camera coupled to a magnifying relay lens. The first hologram used is the *focus positioner*, which generates a double-helix PSF. With this hologram, the rotational orientation of the two lobes provides visual feedback on whether the system is correctly focused. An example is shown in Supplementary Figure 2.

$$\phi(x,y) = \frac{2\pi}{\lambda}\left[f - \sqrt{f^2 + x^2 + y^2}\right] + 2L(x,y)\tan^{-1}\left(\frac{y}{x}\right) \quad \text{(S1.1)}$$

$$L(x,y) = \min\left(1 + \left\lfloor\frac{2N}{D}\sqrt{x^2+y^2}\right\rfloor, N\right), \qquad 0 \leq \sqrt{x^2+y^2} \leq D/2$$

The phase pattern for this hologram is constructed by summing the phase of a double-helix PSF with that of a focusing lens (as shown in Equation S1.1). The double-helix phase contains a factor $L(x,y)$ that represents a stair-case function so that $L(r) = 1,2,3,\ldots,N$ for $0 \leq r < D/2N, D/2N \leq r < D/N, D/N \leq r < 3D/2N, \ldots, (N-1)D/2N \leq r \leq D/2$ where $r = \sqrt{x^2+y^2}$, $D$ is the aperture diameter, and $N \in \mathbb{Z}$ is the number of steps sub-divisions or rings along the diameter. The sum of these phases results in a compound phase mask on the ONN wafer that gives information on alignment at the focal plane. When illuminated, this mask produces two distinct lobes that rotate as a function of axial displacement. We designed the double helix phase such that, at the nominal imaging distance of 2 mm (the ONN working distance) the two lobes are vertically aligned (i.e., 0° rotation with respect to the vertical axis). As shown in Supplementary Figure 2, any deviation from this position, such as a shift of ±50 μm, introduces a clear rotational signature in the PSF. This sensitivity enables precise focusing and fine-tuning of the ONN position along the optical axis when in conjunction with typical micrometer translation stages.

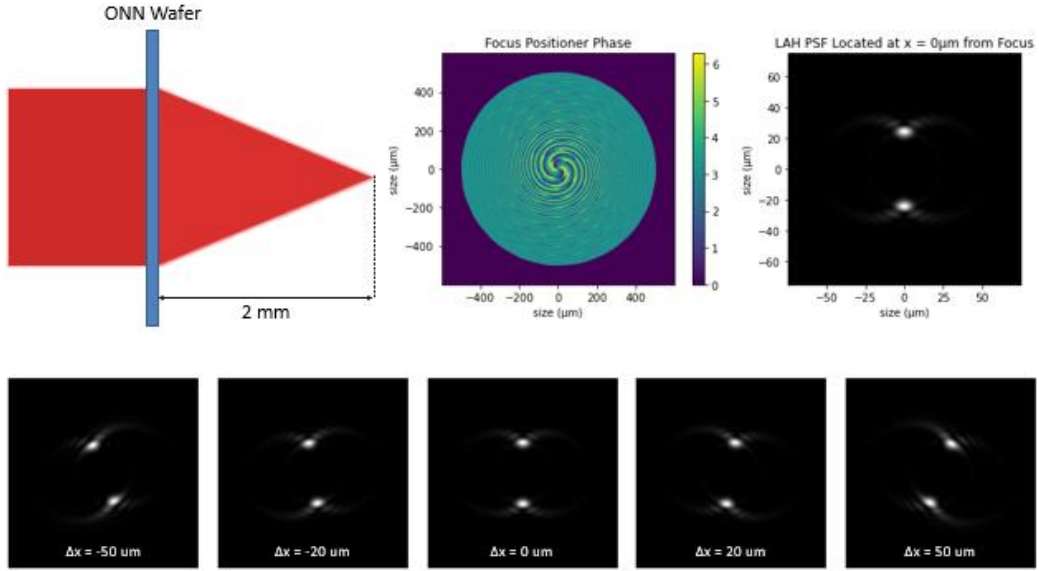

**Supplementary Fig. 2 | Rotational performance of Focus Positioner Hologram.** The focus positioner hologram is used to ensure that the ONN wafer is placed at the correct axial distance from the imaging plane. The rotation of the PSF is used to determine the distance from the focus, where two vertical lobes correspond to perfect positioning, and small deviations from the center corresponding to defocus in each direction

### S1.2 Transverse Alignment Holograms

Once the ONN is properly aligned along the optical axis with respect to the imaging system, the next step is to correct for transverse and longitudinal misalignments. To facilitate this, we implemented *transverse alignment holograms (TAHs)* based on the holointerferometric phase design method described by Ghahremani et al.[2], which enables sub-nanometer alignment between two metasurfaces. These holograms operate in pairs, producing an interferometric pattern at the sensor plane. When perfectly aligned, symmetrical pattern appears. However, deviations in the relative position of the ONN and focusing metasurface cause noticeable distortions in the intensity profile, making the system highly sensitive to small misalignments.

To construct the phase masks, we defined concentric inner and outer rings on both the FMS and ONN wafers (see Supplementary Figure 3). Note that z is the distance from the first surface to the next (50mm in our case); the phase distribution is as follows:

- *FMS Inner Ring*: Focusing phase corresponding to a propagation distance of $\frac{z}{4}$
- *FMS Outer Ring*: Focusing phase corresponding to a propagation distance of $\frac{3z}{4}$
- *ONN Inner Ring*: Focusing phase corresponding to a propagation distance of $\frac{z}{4}$, offset by $\pi$
- *ONN Outer Ring*: Focusing phase corresponding to a propagation distance of $\frac{3z}{4}$
- *Entire ONN Mask*: Add Focusing phase of propagation distance $f = 2mm$ (distance from ONN to sensor)

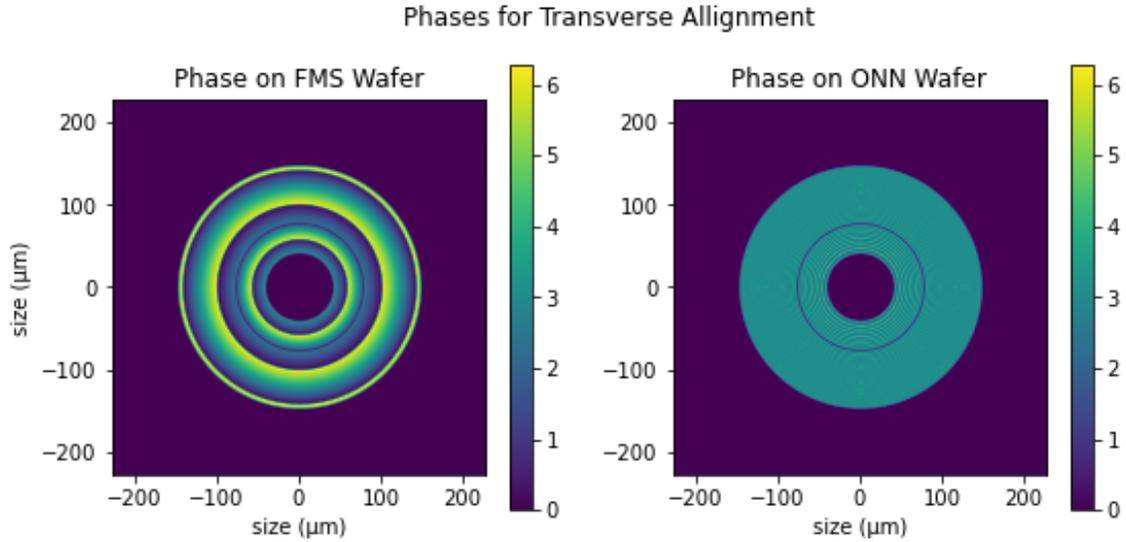

**Supplementary Fig. 3 | Phase for the Transverse Alignment hologram.** The transverse alignment hologram has a phase corresponding to the phase above with the equation shown in Eq.1 at various focusing distances. Small movements transversely leads to discrepancies for the intensity, where the main lobe shifted, as shown in Supplementary Fig.4

The combination of these phase profiles ensures that constructive and destructive interference occurs at specific spatial regions, depending on the alignment quality. A visualization of the resulting interferometric patterns is shown in Supplementary Figure 4. To quantify sensitivity, cross-sectional slices were taken along the central axis of the intensity distribution at various misalignment offsets.

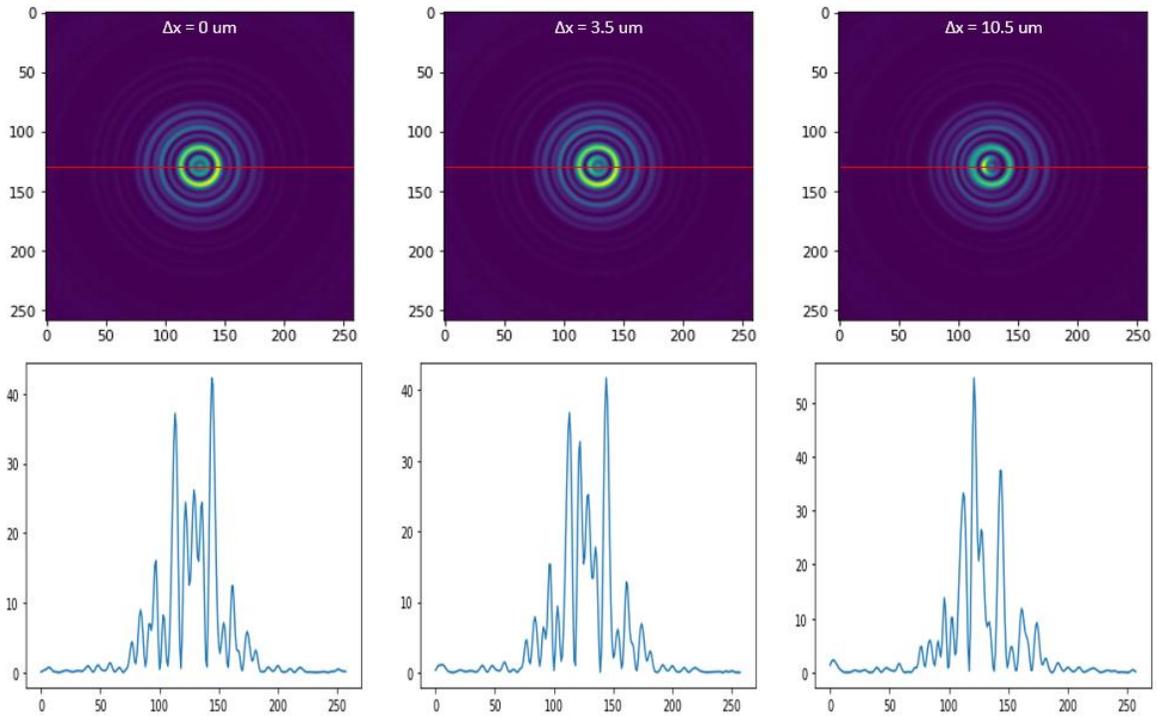

**Supplementary Fig. 4 | Cross-Sectional Intensity Plots for Transverse Alignment.** Based off the phase shown in supplementary Fig. 3, the intensity and shape of the hologram and shift, leading to higher intensity at the outer rings. The hologram enables precision of placement down to the 10 micron level.

Although lateral displacements on the order of 3 μm may not produce visually obvious changes, shifts of ~10 μm result in pronounced alterations in the intensity structure, making this method well-suited for coarse and fine alignment alike. In addition, angular misalignment (tilt) can be found by having multiple TAHs on a single wafer at opposite edges. Any tilt in the system is effectively the same as a slight defocus (due to the increased optical path length) and will result in noticeably different intensity patterns between the two holograms. When aligning in the lab, a method of swapping between using rotation, tip, and transverse correction knobs to decouple all of the misalignments.

**S1.3 Longitudinal Alignment Holograms**

The final alignment element used is the *Longitudinal Alignment Hologram* (LAH). Like the focus positioner hologram, the LAH leverages a double-helix point spread function (PSF) to infer axial misalignments within the system. However, the LAH is implemented using two distinct phase masks: one located on the FMS wafer and the other on the ONN wafer. These two components together generate the required phase structure for the double-helix PSF while also encoding alignment information about the relative positioning of the wafers.

The optical layout and phase distributions for this alignment scheme are shown in Supplementary Figure 5. On the FMS wafer, the LAH phase is patterned into a ring that surrounds the central focusing region (used to create PSFs for the ONN), with additional cutouts at the edges to accommodate the TAHs. The ONN wafer contains two LAH phase masks, positioned on the top and bottom portions of the wafer, which are each composed of a double-helix phase profile combined with a focusing phase to the sensor plane. The focusing phase is spatially offset upward or downward based on the mask's position relative to the ONN's central aperture.

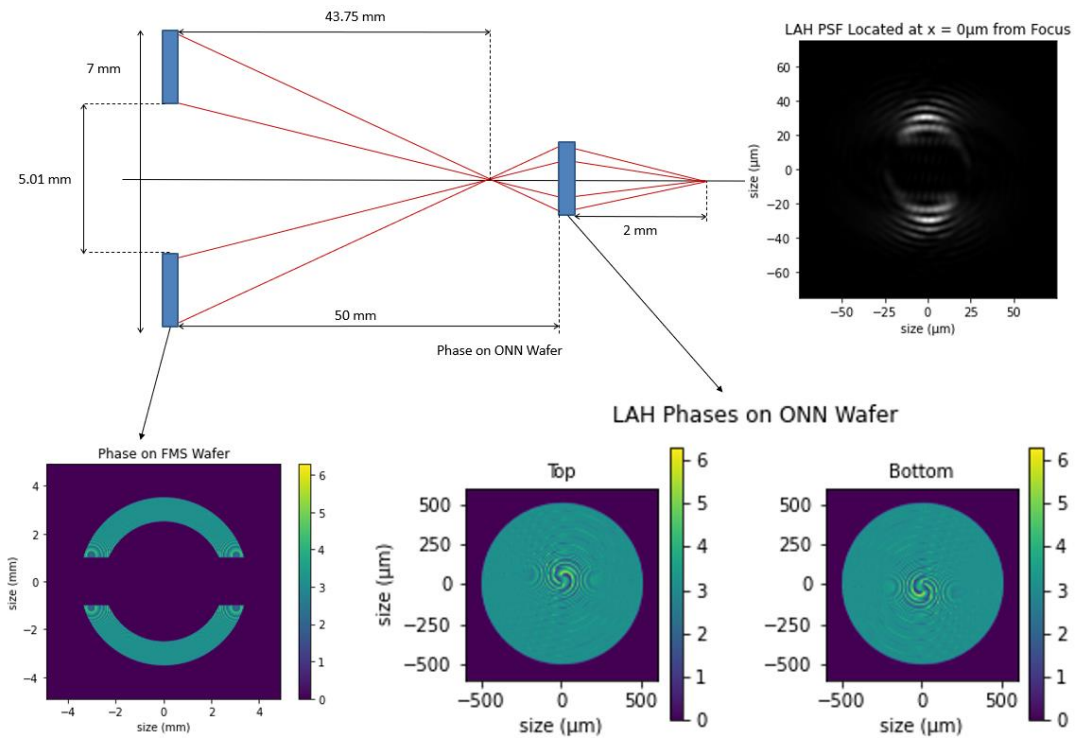

**Supplementary Fig. 5 | Longitudinal Alignment Holograms.** Based on the phases shown for the FMS and Encoder wafer, a double helix PSF can be formed to visualize the placement on the longitudinal axis. The output is similar to the focus positioner; however, it is to describe the location of the FMS in reference to the ONN.

In terms of correction sensitivity, the LAH is less precise than both the focus positioner and the TAH. This reduced sensitivity is due to the F/# in of the system. Double helix PSF rotation is correlated to how fast the system is, and since we are forced to use a small aperture due to the size of the optics in the wavefront sensor, we are forced to have less-effective LAH. Examples of LAH output can be seen in Supplementary Figure 6. For large axial shifts (on the order of hundreds of microns), the resulting changes in the PSF are readily visible and easily corrected through visual inspection. However, for smaller displacements (tens of microns), the PSF variation is minimal and difficult to detect, limiting the LAH's effectiveness for fine axial alignment. As with the TAHs, the use of two LAHs, placed on opposite edges of the ONN wafer, enabling comparisons between the two outputs. By comparing the double-helix outputs from each side, it is possible to detect and correct for residual tip in the system as tip would increase the optical path length for one output.

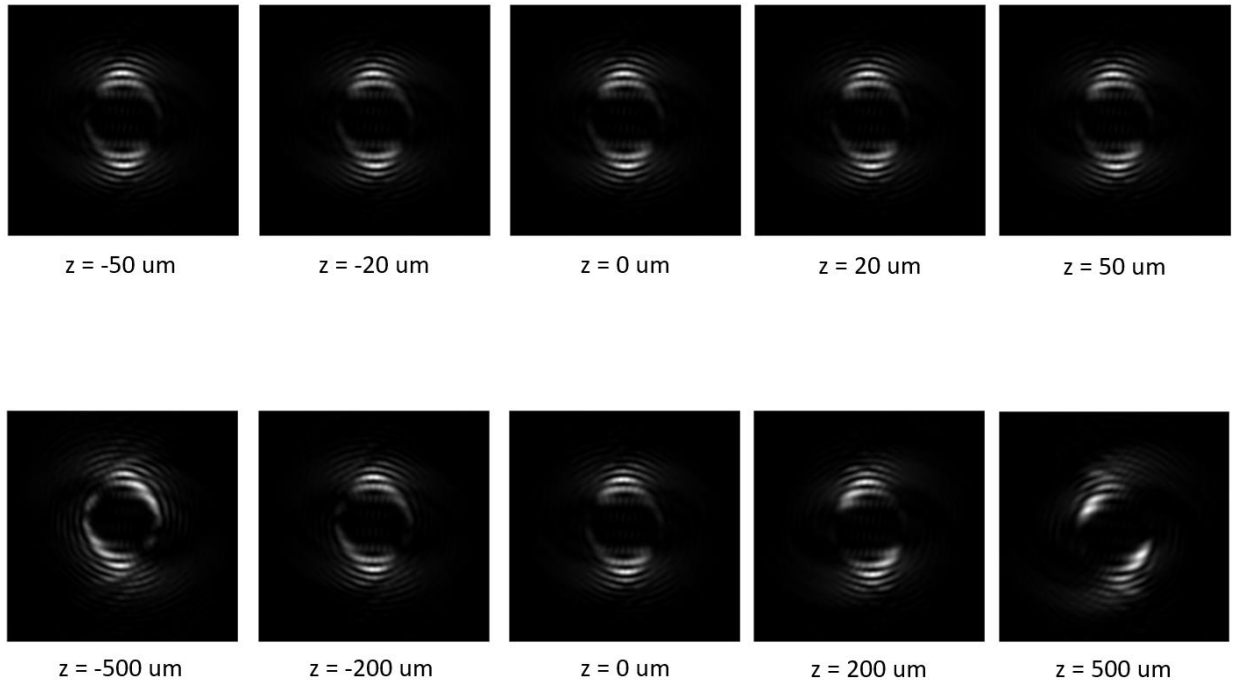

**Supplementary Fig. 6 | Longitudinal Alignment Hologram Performance.** The performance of the LAH is shown in both small-scale and large-scale movements. While the focus positioner shown in Supplementary Fig. 2 can describe the ONN position relative to the focal plane with a couple microns of accuracy, the LAH is much less accurate. This is due to the focal lengths between the FMS and ONN being much larger than they were for the Focus positioner.

Together, the three alignment holograms (focus positioner, TAHs, and LAHs) enable rapid and cost-effective alignment across all six degrees of freedom, with sensitivity ranging from tens to hundreds of microns. For any residual misalignments not resolved through holographic alignment, the ANN is capable of compensating via transfer learning, correcting for minor discrepancies between simulated and experimental inputs.

## S2 Focusing Metasurface and Supercell Design

The focusing metasurface (FMS) employs a supercell approach to generate two distinct point spread functions (PSFs) used as inputs to the optical neural network (ONN). This is achieved through a "checkerboard" pattern in which each white pixel corresponds to the phase required to focus at a specific distance $f$, while each black pixel corresponds to the phase for a slightly longer focal length $f + \Delta f$. Each pixel from both phase profiles is placed in the simulation following the checkerboard pattern to construct the complete phase mask for the focusing metasurface. This method also effectively doubles the spatial resolution of the FMS. This method was specifically chosen as adding the contributions of the focusing phase profile and defocused phase profile do not result in two laterally shifted PSFs. An illustration of this arrangement is provided in Supplementary Fig. 7 below.

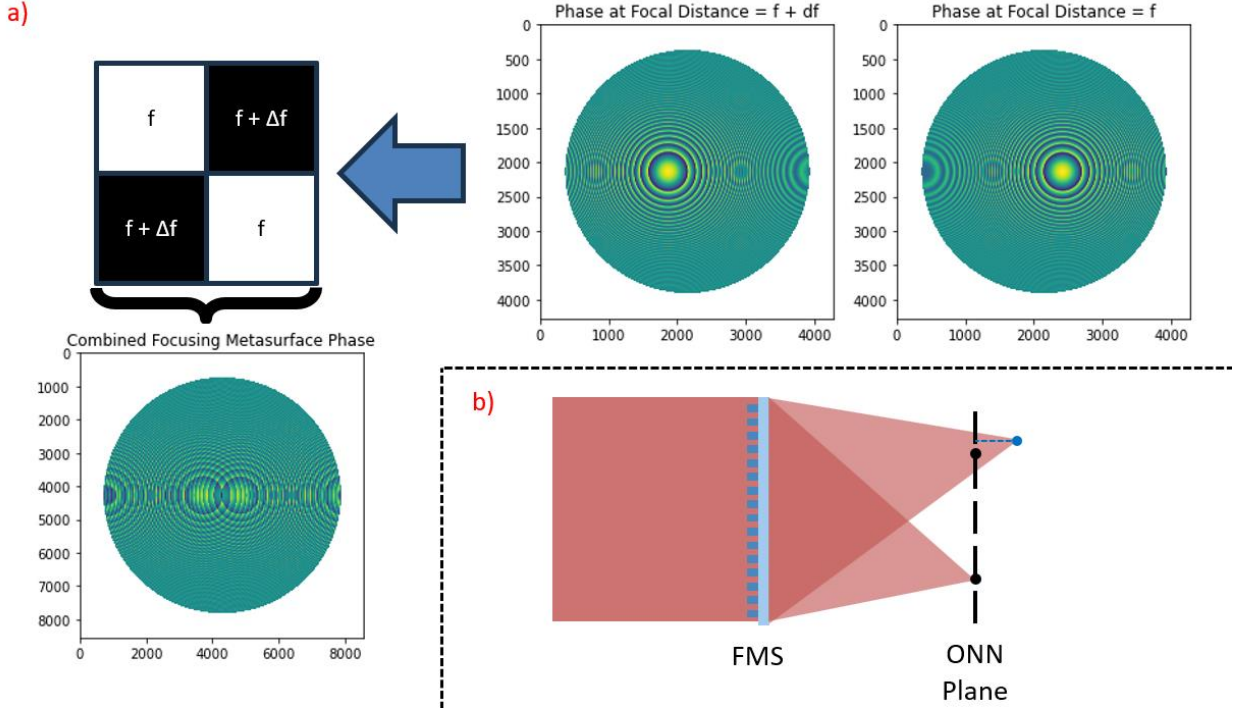

**Supplementary Fig. 7 | Design of the FMS.** The focusing metasurface was created through spatial multiplexing of the unit cells. Two phases were created, one for perfect focus at the input plane and another for a slight defocus. a) The spatial multiplexing process used to create the phase necessary for the focusing metasurface. b) The position of the PSFs with reference to input ONN plane. A slight lateral shift was necessary to ensure the proper spacing between the two PSFs (as shown by the blue line)

The phase profiles for both the focused and defocused components of the metasurface are generated using the standard focusing phase equation (Equation S2.1). Note that $n$ and $m$ are the discretized coordinates used to oscillate between $f_0 = 50\ mm$ and $f_1 = 52.4\ mm$ and the horizontal offset, $x_0$, depending on the summation of $n$ and $m$ being even or odd (Equation S2.1).

$$\varphi(m,n) = \frac{2\pi}{\lambda}\left[f_{\chi(n,m)} - \sqrt{f_{\chi(n,m)}^2 + (m\Delta x - (-1)^{\chi(n,m)}x_0)^2 + (n\Delta y)^2}\right] \quad (S2.1)$$

$$\chi(n,m) = \begin{cases} 0, & n+m \equiv 0\ (mod\ 2) \\ 1, & n+m \not\equiv 0\ (mod\ 2) \end{cases}$$

$$m \in \mathbb{Z}, \quad -\frac{D}{2\Delta x} \leq m \leq \frac{D}{2\Delta x}$$

$$n \in \mathbb{Z}, \quad -\frac{D}{2\Delta y} \leq n \leq \frac{D}{2\Delta y}$$

Here:
- $n$ and $m$ are the discretized coordinates
- $f_{\chi(n,m)}$ is the focal length of $f_0$ or $f_1$ if $n+m$ even or odd, respectively

- $\lambda$ is the operational wavelength
- $D$ is the lens diameter
- $x_0$ is the horizontal offset of 374.5 μm
- $x$ and $y$ are spatial locations with $\Delta x$ and $\Delta y$ being step sizes

Since the ONN is designed with a lateral spacing of 749 μm between PSFs, the generated PSFs must be aligned accordingly. However, due to the defocus applied to one of the PSFs, its horizontal position at the ONN plane deviates slightly from the ideal 749 μm separation, as shown in Supplementary Fig.7b. As such, a small correction ($x_0$) was applied to the x-coordinate in the phase equation to compensate for this offset and ensure accurate alignment of the PSFs at the ONN input plane.

## S3 Deformable Mirror Model

To simulate a deformable mirror during training, we employed an influence function to model the effect of each actuator on the mirror surface. While traditional approaches often involve complex mathematics, Gaussian influence functions have proven to be among the simplest and most effective, achieving an $R^2$ value of 0.977 [3]. Subsequent research has introduced the modified Gaussian influence function, which improves accuracy and achieves an $R^2$ value of 0.999 [4]. This makes it a highly accurate model for actuator behavior. Additionally, the function is well-suited for neural network-based training, helping reduce both training time and computational complexity during phase reconstruction.

The modified Gaussian influence function is defined as:

$$I(r,\theta) = \exp\left( ln(\omega) \cdot \left\{ \frac{r \cdot [1 + \lambda \cdot cos(6\theta)]}{d} \right\}^\alpha \right) + \beta \cdot \exp\left(-\frac{r-d}{\gamma^2}\right) \quad (1)$$

In this equation:

- r and θ are the radial and angular coordinates,
- λ is the azimuthal modulation amplitude,
- d is the mean actuator spacing,
- ω is the inter-actuator coupling constant,
- α is the Gaussian power index,
- β is the radial correction amplitude, and
- γ is the radial correction width.

The parameters β and γ typically range from 0.01–0.06 and 3–10, respectively, depending on the specific deformable mirror being modeled [3]. In our case, which uses a hexagonal actuator layout, Huang et al. have demonstrated that γ=6 and β=0.02 yield accurate results [4].

Our hybrid system is optimized for a 5 mm deformable mirror. The system outputs control values in the range [−1,1] where:

- 1 corresponds to the maximum positive actuator stroke
- 0 corresponds to a flat, inactive actuator, and
- -1 corresponds to the maximum negative actuator stroke.

Illustrations of single-actuator activation, all-actuators-on configurations, and a representative wavefront correction are provided in Supplementary Figure 8.

To reconstruct the correction phase from the neural network outputs, we superimpose the contributions of all actuators to form a complete phase profile. To enhance edge behavior and better replicate real-world conditions, we also include actuators positioned outside the mirror's aperture. These peripheral actuators help enforce smooth phase transitions at the boundaries during simulation and increase correction at the boundary.

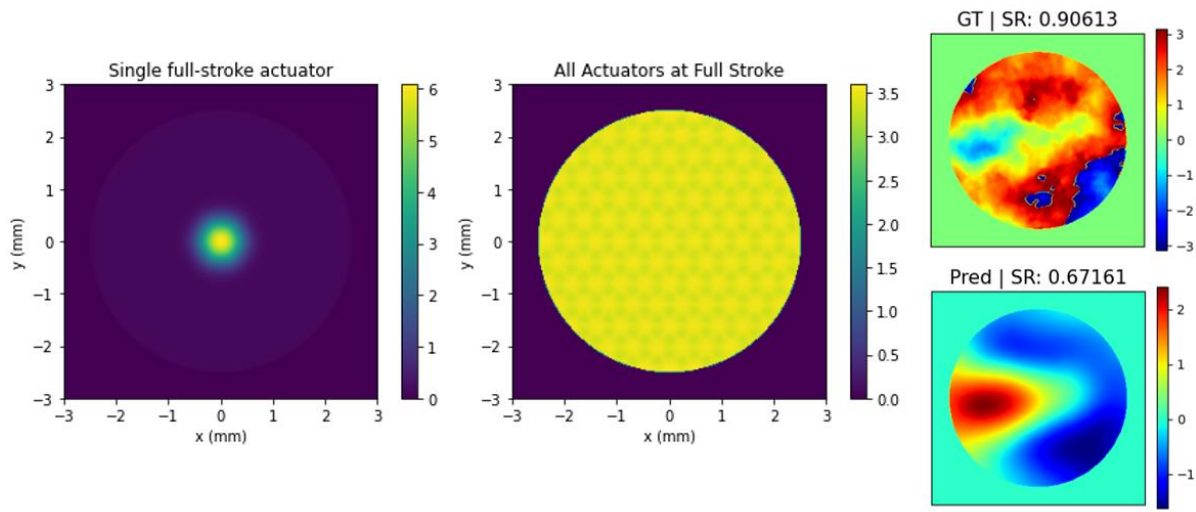

**Supplementary Fig. 8 | Deformable Mirror Modelling.** The deformable mirror was modeled with gaussian influence function at each actuator and the value of each actuator was made based off the relative stroke for each actuator (as described above). We can see the phase imparted by the DM with all actuators at max as well as the correction of the DM on a wavefront.

### S4 Latency Measurements

We measured the forward pass runtime of all MLP models in C++ by using the LibTorch library to capture the models written in Pytorch via compilation. Loading and running the models on C++ allowed for more precise measurements of their execution times. Table 1 shows the measured runtimes $\tau$ averaged over 1000 runs done on an AMD Ryzen Threadripper PRO 5955WX CPU with 16 cores and an NVIDIA RTX A6000 Ada Generation GPU. We note that for the smaller models, the CPU offers faster performance, while bulkier models (like the 2x4096) highly benefit from the GPU parallelization. This offers a promising solution for improving our sensor's accuracy while maintaining low latencies for high-speed operation.

**Table 1.** Latency measurements for all models on CPU and GPU.

| Input Size | Hidden Layers | Hidden Nodes | $\tau_{CPU}$ (μm) | $\tau_{GPU}$ (μm) |
|---|---|---|---|---|
| 3x3 | 1 | 16 | 10.539 | 22.51502 |
| 3x3 | 1 | 128 | 11.717 | 22.36601 |
| 3x3 | 1 | 1024 | 24.46298 | 22.62899 |
| 3x3 | 2 | 4096 | 1478.852 | 37.69105 |
| 6x6 | 1 | 1024 | 28.25801 | 22.89 |
| 8x8 | 1 | 1024 | 34.08502 | 25.798 |
| 24x24 | 1 | 1024 | 82.428 | 25.798 |
| 168x168 | 1 | 1024 | 5344.735 | 269.8917 |